\documentclass[epjST]{svjour}

\newcommand{\spin}[1]{\sigma_{#1}} 
\newcommand{\nn}[2]{\left<#1,#2\right>} 
\newcommand{\nnn}[2]{\left<\left<#1,#2\right>\right>} 
\newcommand{\plaq}[4]{\left[#1,#2,#3,#4\right]} 
\newcommand{\nnsum}{\sum\limits_{\nn{i}{j}}\spin{i}\spin{j}} 
\newcommand{\nnnsum}{\sum\limits_{\nnn{i}{j}}\spin{i}\spin{j}} 
\newcommand{\plaqsum}{\sum\limits_{\plaq{i}{j}{k}{l}}%
	\spin{i}\spin{j}\spin{k}\spin{l}}

\mathchardef\mhyphen="2D

\def\be{\begin{equation}}
\def\ee{\end{equation}}
\def\bea{\begin{eqnarray}}
\def\eea{\end{eqnarray}}

\usepackage{amsmath,amssymb}
\usepackage{dcolumn}
\usepackage{braket}
\usepackage{hhline}
\usepackage{graphicx}
\usepackage{mathtools}
\usepackage{xfrac}
\usepackage{paralist}
\usepackage{enumitem}
\usepackage[normalem]{ulem}
\usepackage{color}
\definecolor{darkblue}{rgb}{0.,0.,0.4}
\definecolor{darkred}{rgb}{0.5,0.,0.}
\usepackage{booktabs}
\usepackage{pdflscape}

\newcommand{\betainfall}{0.551\,334(8)}

\usepackage{colortbl}
\definecolor{lightgrey}{rgb}{0.9,0.9,0.9}
\definecolor{grey}{rgb}{0.7,0.7,0.7}

\begin{document}
\title{Plaquette Ising models, degeneracy and scaling}
\author{
  Desmond A.\ Johnston\inst{1}\fnmsep\thanks{\email{D.A.Johnston@hw.ac.uk}} \and
  Marco Mueller\inst{2}\fnmsep\thanks{\email{Marco.Mueller@itp.uni-leipzig.de}} \and
  Wolfhard Janke\inst{2}\fnmsep\thanks{\email{Wolfhard.Janke@itp.uni-leipzig.de}}
}
\institute{Department of Mathematics, 
	School of Mathematical and Computer Sciences,
	Heriot-Watt University, Riccarton, Edinburgh EH14 4AS, Scotland
  \and Institut f\"ur Theoretische Physik, Universit\"at Leipzig,
  Postfach 100\,920, 04009 Leipzig, Germany}
\abstract{
We review some recent investigations of the $3d$ plaquette Ising model. This
displays a strong first-order phase transition with unusual scaling properties
due to the size-dependent degeneracy of the low-temperature phase. In
particular, the leading scaling correction is modified from the usual inverse
volume behaviour $\propto 1/L^3$ to $1/L^2$.  The degeneracy also has
implications for the magnetic order in the model which has an intermediate
nature between local and global order and gives rise to novel fracton
topological defects in a related quantum Hamiltonian.
} 
\maketitle
\section{Introduction}
\label{intro}
The Ising model \cite{Ising} with ferromagnetic nearest-neighbour interactions
serves as a paradigm for ferromagnetic phase transitions and is almost
certainly the most studied single model in statistical mechanics
\cite{mccoy1973}. In general, outside the field of disordered systems Ising
models with multispin interactions have been less explored
\cite{Lip1,JLMreview}, both because such interactions are less common in real
materials and also because in many cases such models display first-order
transitions. These might be regarded as a priori less promising subjects for
numerical investigation, although the general scaling theory for first-order
transitions, initiated in \cite{pioneer} and developed further in
\cite{furtherd} and \cite{rigorous-fss}, is now generally well-understood. In
disordered systems, of course, multispin interaction Hamiltonians  play an
important role in the random first-order transition (RFOT) theory of spin
glasses \cite{RFOT}.

Here, we remain within the ambit of purely ferromagnetic couplings and review
some recent work on one of the simplest multispin interaction models, a $3d$
plaquette Ising Hamiltonian with the spins $\sigma=\pm 1$ sited at the vertices
of a $3d$ cubic lattice,
\begin{equation}
\label{e2k}
H =  -  \frac{1}{2} \sum_{[i,j,k,l]}\sigma_{i} \sigma_{j}\sigma_{k} \sigma_{l} \; .
\end{equation}
The four-spin interactions in the Hamiltonian are between  spins situated on
the corners of the plaquettes that comprise each cube.

The purely plaquette Hamiltonian of Eq.~(\ref{e2k}) can be thought of as the
limiting case, for $\kappa \to 0$,  of a one-parameter family of $3d$
gonihedric Ising Hamiltonians \cite{savvidy1,savvidy2}. These contain in
general nearest-neighbour $\langle i,j \rangle$, next-to-nearest-neighbour
$\langle \langle i,j \rangle \rangle$ and plaquette $[i,j,k,l]$ interactions,
\begin{equation}
H^\kappa = -2\kappa\nnsum+\frac{\kappa}{2}\nnnsum-\frac{1-\kappa}{2}\plaqsum\;,
\label{ham:goni}
\end{equation} 
where the ratios of the various couplings have been fine-tuned to eliminate the
bare surface tension, which in the case of the standard nearest-neighbour Ising
Hamiltonian is the only term. When regarded as a model of fluctuating surfaces
described by the geometric spin cluster boundaries the intention is to weight
the edge length of spin clusters rather than their boundary area
\cite{Cappi}.

For $\kappa \ne 0$ the $3d$ Hamiltonians $H^\kappa$  display a continuous
transition, possibly with $3d$ Ising exponents which are masked by strong
crossover effects from the nearby tricritical point \cite{Italians}. The $\kappa=0$
plaquette Hamiltonian, on the other hand, has a strong first-order phase
transition \cite{firstorder} and recent high-precision multicanonical
simulations \cite{us_goni} have revealed that it displays non-standard
finite-size scaling properties at this transition.  In the sequel we will
describe how this non-standard scaling is a  consequence of a macroscopic
low-temperature phase degeneracy in the $3d$ plaquette Ising model and also
discuss the implications of this degeneracy for  the nature of the
low-temperature phase and the definition of an order parameter
\cite{me_fukinuke,us_fukinuke}.  Recent work on quantum spin versions of the
plaquette model  has also shown that the symmetries of the plaquette
Hamiltonian and the low-temperature degeneracy are  instrumental in the
appearance of a novel form of topological defect, a fracton
\cite{fracton1,fracton2,fracton_hist}, and we sketch their appearance via a
partial gauging construction using  the quantum dual of the plaquette model.

\section{A Curious Symmetry -- Classical Aspects}

\begin{figure}[b]
	\centering
	\includegraphics[width=4cm]{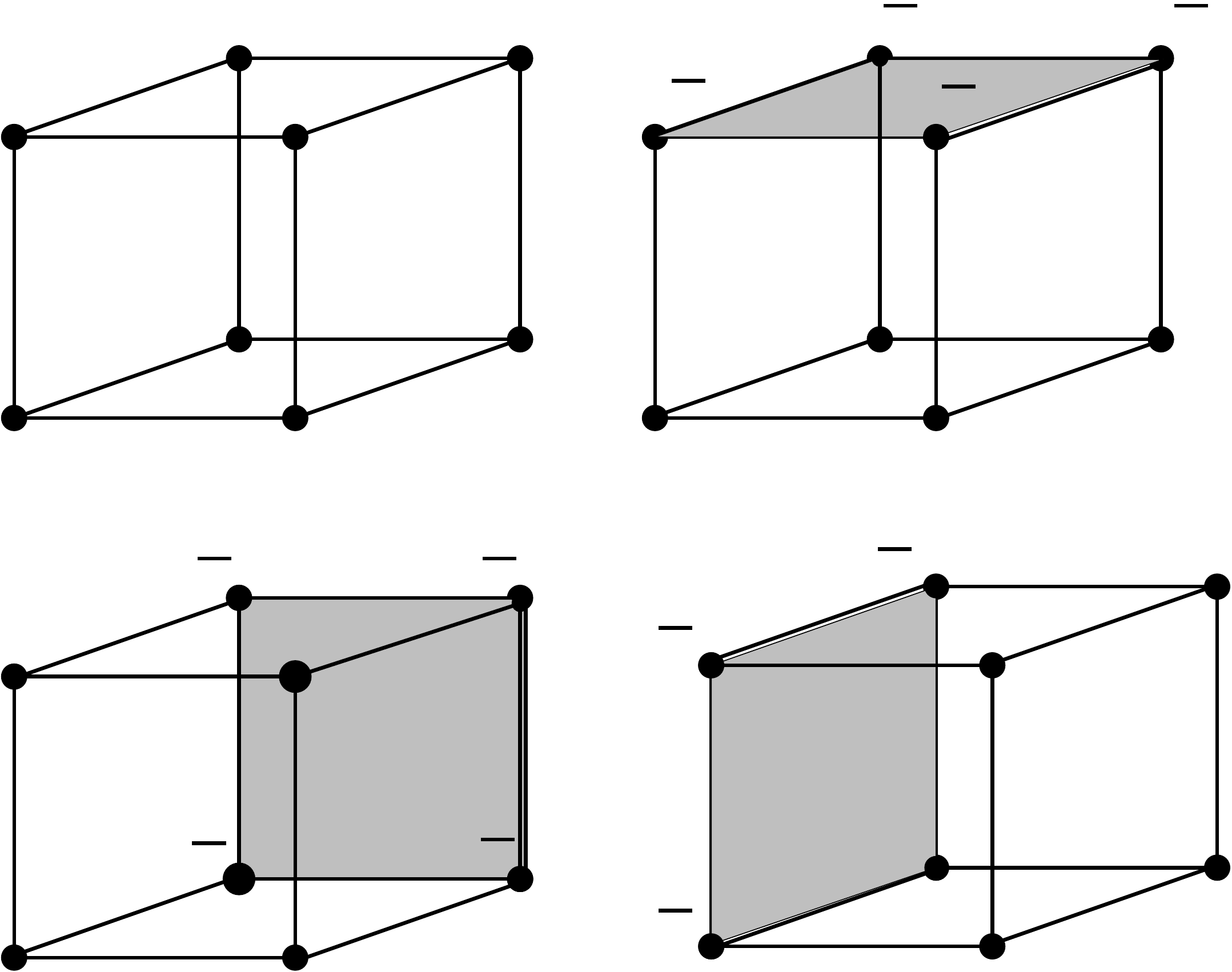}
  \caption{Flipping the value of the Ising spins on a face of a single cube
    does not change its contribution to the energy at $T=0$. All of the
    configurations shown have the same energy.} 
	\label{fig:sketch}
\end{figure}

The Hamiltonians of Eqs.~(\ref{e2k}) and (\ref{ham:goni}) have symmetry
properties which are intermediate between those of a gauge theory and a global
symmetry.  Consideration of how the spin configuration on a cubic lattice can
be built up from the individual cubes \cite{Cappi,malpap1} shows that the spins on
any face(s) of a single cube may be flipped at zero energy cost.
Since an entire ground-state  lattice configuration may be built by stacking
compatibly shaded cubes this means that that there is freedom to flip planes of
spins at zero energy cost, see Fig.~\ref{fig:sketch}. 

\begin{figure}[t]
	\begin{center}
	\includegraphics[width=3.5cm]{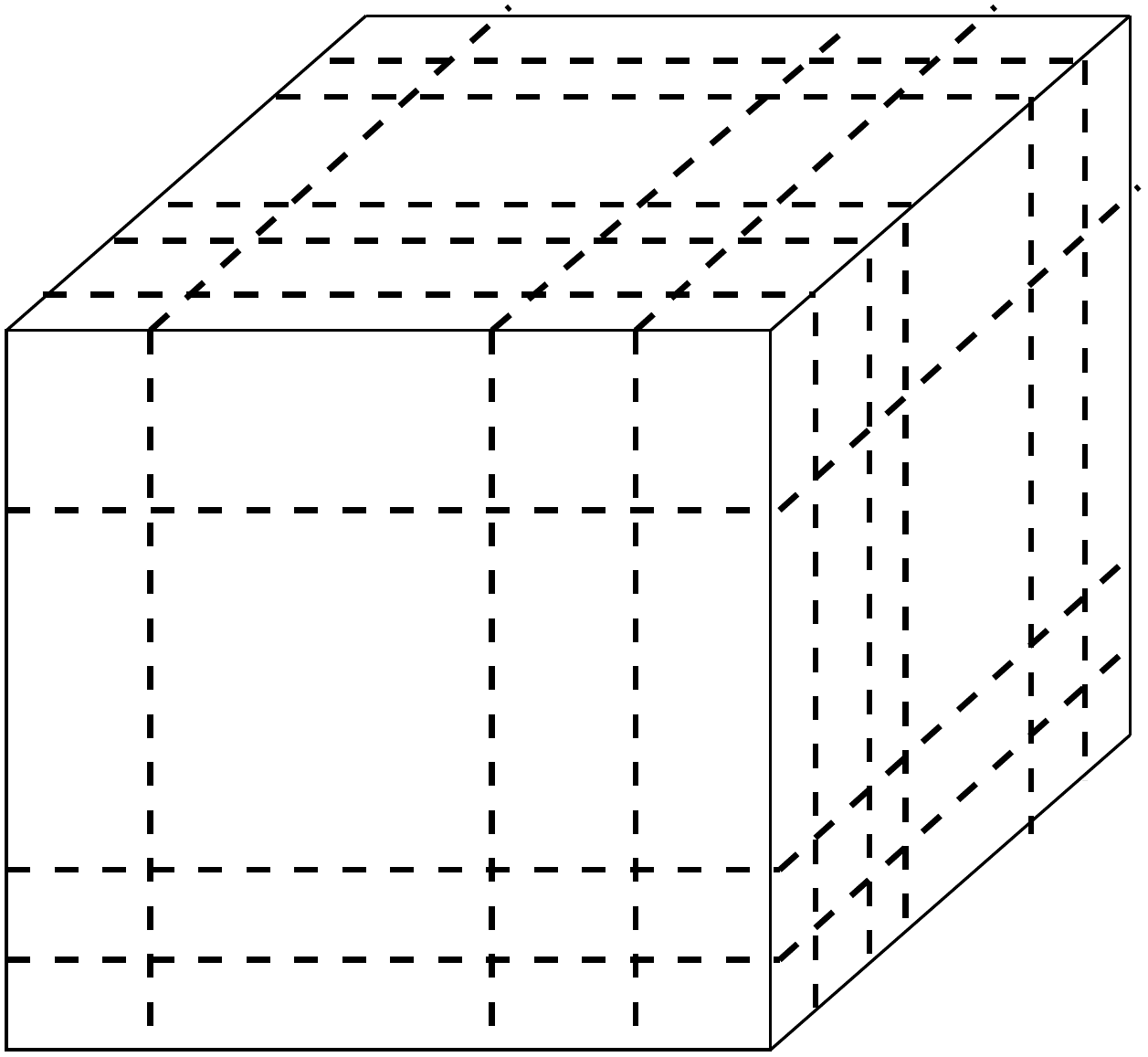}
  \caption{A typical ground state of the $3d$ plaquette Hamiltonian showing
    the edges of the planes of spins that are flipped with respect to a purely
    ferromagnetic ground state dotted. Since {\it any} plane of spins in any of
    the three possible orientations may be flipped in such a configuration, the
    ground-state degeneracy on an $L^3$ lattice is $q=2^{3L}$.} 
	\label{fig:sketch2}
	\end{center}
\end{figure}

When $\kappa \ne 0$ we can see that parallel non-intersecting planes of spins
may be flipped at zero energy cost at zero temperature, giving a $3 \times
2^{2L}$ ground-state degeneracy on an $L^3$ cubic lattice. When $\kappa = 0$,
on the other hand, the constraint on intersections is relaxed so any planes may
be flipped and the ground-state degeneracy becomes $2^{3L}$,
as illustrated in Fig.~\ref{fig:sketch2}.

When $\kappa \ne 0$ the ground-state degeneracy is broken at finite
temperature, as is revealed by a low-temperature expansion
\cite{pietig_wegner}. When $\kappa=0$, however, the $2^{3L}$ degeneracy
persists into the low-temperature phase.
Such macroscopic {ground-state} degeneracy is not uncommon in other systems,
what is rarer is its persistence into the low-temperature phase. The
nearest-neighbour Ising antiferromagnet on a FCC lattice, for instance, also
has a macroscopically degenerate ground-state in which crystal planes of spins
may be flipped, but only six possible  ordered phases survive from these at
finite temperature  \cite{beath}.  

\begin{figure}[b]
	\centering
	\includegraphics[height=4cm]{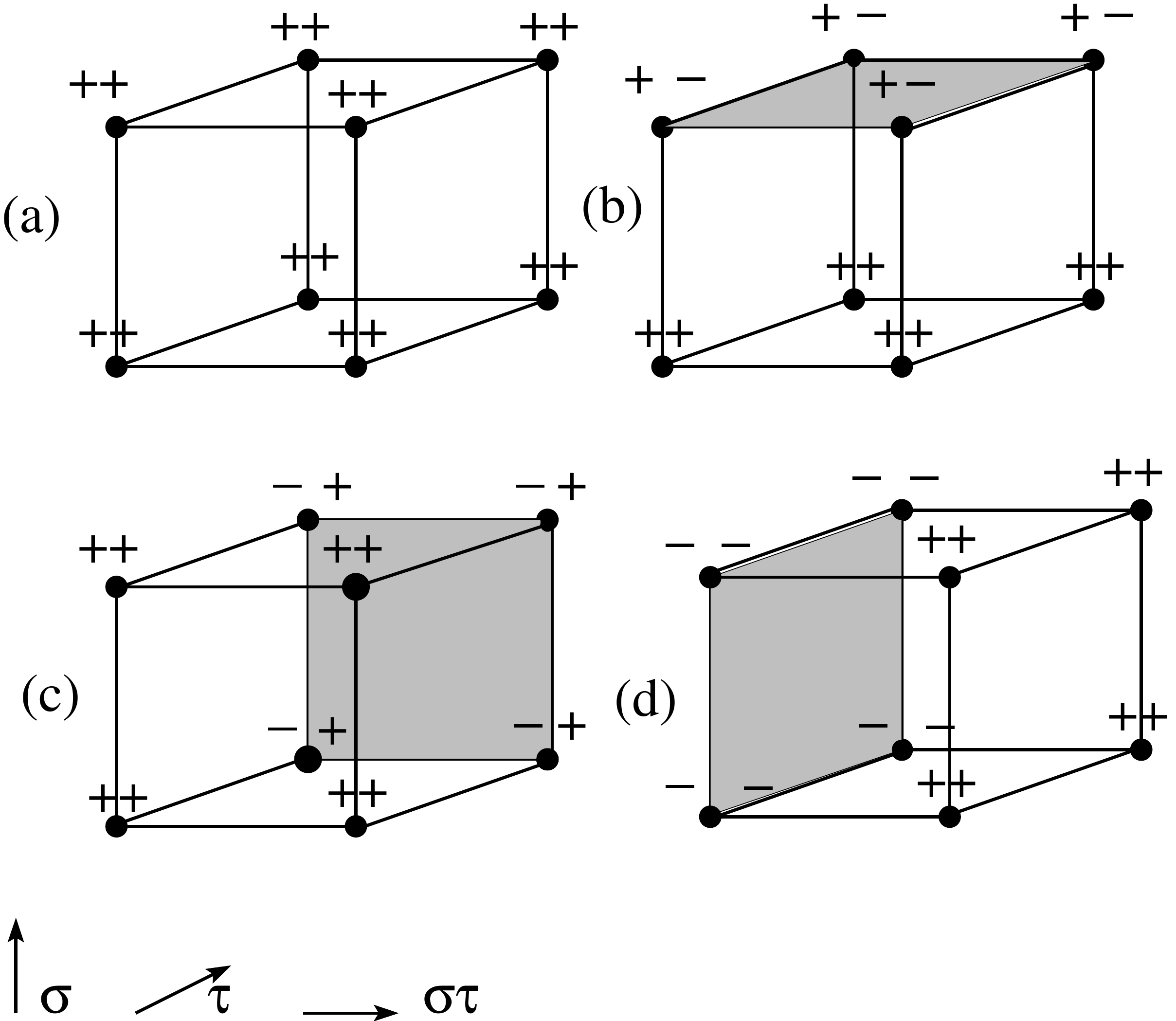}
  \caption{Some possible ground-state spin configurations for the dual model on
    a cube with flipped spins. The $\sigma,\tau$ values are shown at each site,
    as are the directions of the anisotropic interactions between the spins.
    Once again, the entire lattice may be tiled by compatible combinations of
    such cubes.}
	\label{ground} 
\end{figure}

The dual of the plaquette Hamiltonian is an anisotropically coupled
Ashkin-Teller model
\begin{equation} 
	H_{\rm dual} = - \frac{1}{2} \sum_{ \langle ij \rangle} \sigma_{i}  \sigma_{j} 
	- \frac{1}{2}  \sum_{ \langle ik \rangle } \tau_{i}  \tau_{k} 
	-  \frac{1}{2} \sum_{\langle jk \rangle} \sigma_{j} \sigma_{k} \tau_{j}  \tau_{k} \, ,
	\label{dual2}
\end{equation}
with two flavours of Ising spins $\sigma, \tau$ \cite{malpap_dual} and
possesses the same ground-state degeneracy as the original plaquette
Hamiltonian, since it is still possible to flip planes of spins at zero energy
cost, see Fig.~\ref{ground}.
Although it has not been confirmed with a low-temperature expansion in the
manner of the plaquette Hamiltonian, the finite-size scaling properties at the
first-order transition in the dual model show that the macroscopic degeneracy
persists into the low-temperature phase there too, as we discuss below. 

In the rest of this  review we will describe the consequences of the
macroscopic low-temperature phase degeneracy in the $3d$ plaquette Ising model
(and its dual) for the finite-size scaling at the first-order phase transition
and the implications for the definition of an order parameter. We then outline
the role played by the spin-flip symmetry in enabling the appearance of fracton
topological defects in a Hamiltonian related to the dual of the quantum version
of the model.

\section{Standard First-Order Scaling}
\label{sec:1}

We can obtain the basic finite-size scaling laws for  first-order phase
transitions by using a simple heuristic two-phase model, as described in
\cite{twophasemodel}.  In this, fluctuations within phases are ignored and we
treat the phase transition as a sharp jump between the phases.  A system thus
spends a fraction $W_{\rm o}$ of the total time in a simulation in one of the
$q$ ordered phases and a fraction $W_{\rm d} = 1 - W_{\rm o}$ in the disordered
phase, with corresponding energies $\hat{e}_{\rm o}$ and $\hat{e}_{\rm d}$. In
these expressions, the hat denotes quantities evaluated at the inverse
transition temperature of the {infinite} system, $\beta^\infty$.  The
expectation values of the energy moments in the model  are then given by
weighted averages over the ordered and disordered phases
\begin{equation} 
\left<e^n\right> =
W_{\rm o}\hat{e}_{\rm o}^n + (1-W_{\rm o})\hat{e}_{\rm d}^n \; .
\end{equation}
The specific heat $C_V(\beta, L) = -\beta^2\partial e(\beta, L)/\partial\beta$
in such an approximation is given by
\begin{equation}
C_V(\beta, L) = L^d\beta^2\left(\left< e^2\right> - \left< e \right>^2\right) =
L^d\beta^2 W_{\rm o}(1-W_{\rm o})\Delta \hat{e}^2
\label{eq:specheat}
\end{equation}
where $\Delta \hat{e} = \hat{e}_{\rm d} - \hat{e}_{\rm o}$. Maximizing this
with respect to $W_o$ gives
\begin{equation}
C_V^{\rm max} = L^d (\beta^\infty\Delta \hat{e}/2)^2
\label{eq:specheatmax}
\end{equation}
at the inverse temperature $\beta^{C_V^{\rm max}}(L)$ for which $W_{\rm o} =
W_{\rm d} = 0.5$. 
%
%
We can take the probability of being in any of the ordered states or the
disordered state to be given by the standard Boltzmann factors
\begin{equation}   
p_{\rm o}\propto e^{-\beta L^d \hat{f}_{\rm o}}\mbox{ and } p_{\rm d} \propto
e^{-\beta L^d \hat{f}_{\rm d}} \; ,
\end{equation}
where $\hat{f}_{\rm o}, \hat{f}_{\rm d}$ are the free energy densities of the
states.  The fraction of time spent in the $q$ ordered states is proportional
to $q p_{\rm o}$ and the  fraction of time spent in the disordered state is
proportional to $p_{\rm d}$, so the ratio of these is given by 
$W_{\rm o}/W_{\rm d} \simeq q e^{- L^d \beta \hat{f}_{\rm o}}/ e^{-\beta L^d \hat{f}_{\rm d}}$ 
up to exponentially small corrections in $L$ \cite{rigorous-fss,borgs-janke}.
Taking the logarithm of this ratio gives 
\begin{equation}   
\ln (W_{\rm o}/W_{\rm d}) \simeq \ln q + L^d\beta(\hat{f}_{\rm d}
- \hat{f}_{\rm o}) \; .
\end{equation} 
Since $W_{\rm o} = W_{\rm d}$ at the finite-size specific-heat maximum,
we find by expansion around $\beta^\infty$ that
\begin{equation}
0 = \ln q + L^d\Delta \hat{e}(\beta -
\beta^\infty) + \dots 
\end{equation}
where $\Delta \hat{e} =  \hat{e}_{\rm d} -  \hat{e}_{\rm o}$, which can be solved for the
location of finite-size peak of the specific heat:
\begin{equation}
\beta^{C_V^{\rm max}}(L) = \beta^\infty - \frac{\ln q}{L^{d}\Delta\hat{e}} +
\dots
\label{eq:fss:beta:specheat}
\end{equation}
Analogous calculations for the energetic Binder parameter 
\begin{equation}
B(\beta, L) = 1 - \frac{\langle e^4\rangle}{3\langle e^2 \rangle^2} \;
\label{eq:binder}
\end{equation}
give~\cite{twophasemodel,lee-kosterlitz}
\begin{equation}
\beta^{B^{\rm min}}(L) = \beta^\infty - \frac{\ln(q\hat{e}^2_{\rm
		o}/\hat{e}^2_{\rm d})}{L^d\Delta \hat{e}}  + \dots 
\label{eq:fss:beta:binder}
\end{equation}
for the location $\beta^{B^{\rm min}}(L)$ of its minimum, which is an
alternative estimator for the finite-size transition point.

The key features of finite-size scaling for first-order transitions are already
exposed in this simple model. In Eq.~(\ref{eq:specheatmax}) we can see that the
specific heat grows with the system volume $L^d$ and  the finite-size values of
the estimators for the (inverse) transition temperatures in
Eq.~(\ref{eq:fss:beta:specheat}) and Eq.~(\ref{eq:fss:beta:binder})  are shifted by
$1/L^d$. This behaviour is generic for finite-size scaling at first-order
transitions for systems with periodic boundary conditions and it is confirmed
by more sophisticated analyses using Pirogov-Sinai theory in cases where a
contour expansion exists, such as for the $q$-state Potts model
\cite{rigorous-fss}. 

We can extend the discussion within the framework of the heuristic model to
encompass non-periodic boundary conditions by allowing surface free-energy
density terms,  $\hat{s}_{\rm o}$ and $\hat{s}_{\rm d}$ for the ordered and disordered
phases, respectively, in the Boltzmann factors. Note that $\hat{s}_{\rm o}$ and
$\hat{s}_{\rm d}$ will in general not be equal since we have phase coexistence at the
first-order transition point.  With such boundary surface energy terms
contributing, the Boltzmann factors become
\begin{equation}
  p_{\rm o}\propto   e^{-\beta (L^d \hat{f}_{\rm o} + {L^{d-1}} \hat{s}_{\rm o})} \mbox{ and } p_{\rm d }\propto   e^{-\beta ( L^d \hat{f}_{\rm d} + { L^{d-1} } \hat{s}_{\rm d})} 
\end{equation}
and we can see that we might expect $1/L$ leading corrections in such
circumstances, which is again confirmed by more sophisticated analytical
arguments and in simulations~\cite{fssfbc}. 

\section{Non-Standard First-Order Scaling}
\label{sec:2}

\begin{figure}[h!]
	\begin{center} 
		\includegraphics[width=0.66\textwidth]{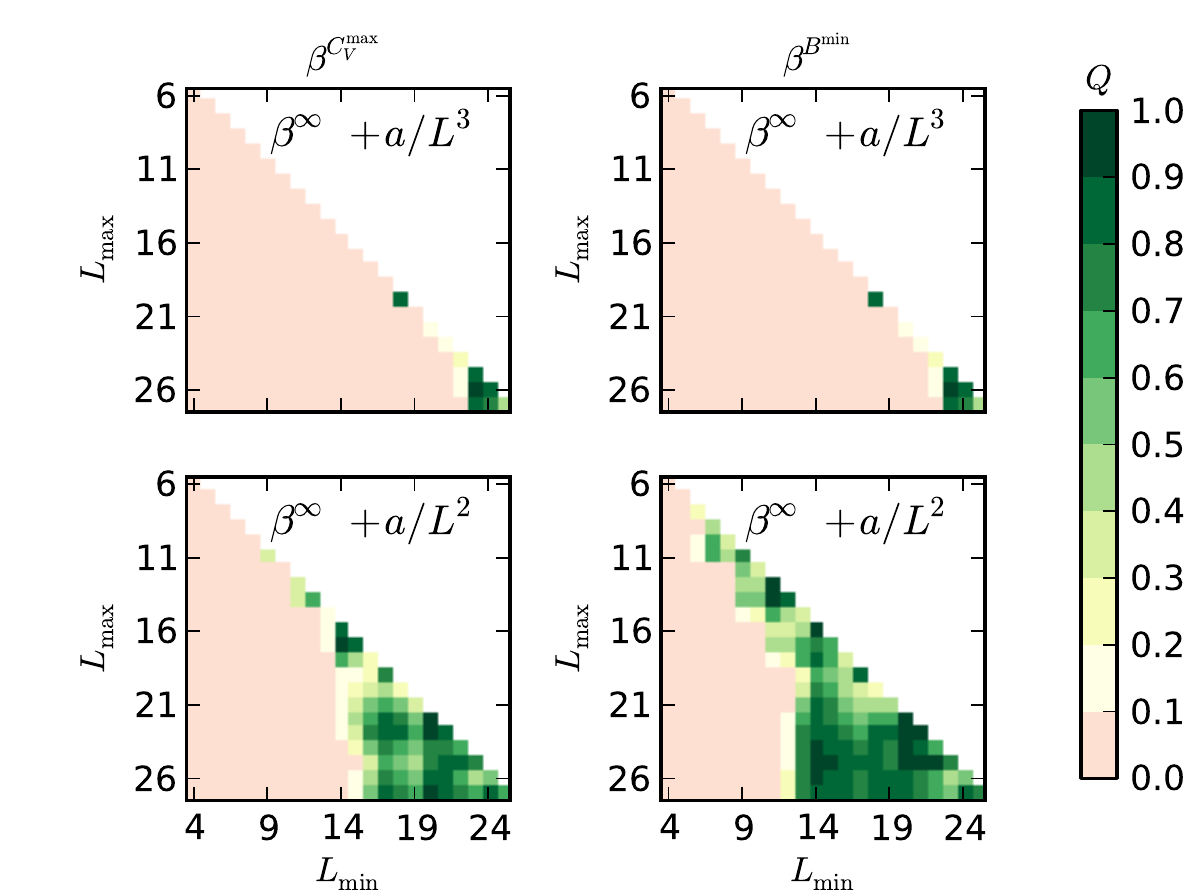}
		\caption{Plot of the goodness-of-fit parameter $Q$ for fits on the
			extremal locations of the specific heat, $\beta^{C_V^{\rm max}}$, and
			Binder's energy cumulant, $\beta^{B^{\rm min}}$, of the $3d$ plaquette Ising model
			for different fitting ranges
      \mbox{$L_{\rm min}$ -- $L_{\rm max}$}. Green regions indicate acceptable
      fits. Upper row: Standard $1/L^3$ finite-size scaling ansatz. Lower row:
      $1/L^2$ finite-size scaling.} 
		\label{fig:fitting_quality} 
	\end{center}
\end{figure}

In principle the $3d$ plaquette Ising model should provide an ideal arena for
exploring the finer points of finite-size scaling at first-order transitions,
since it has a relatively simple Hamiltonian and displays a strong first-order
transition. However, high precision multicanonical simulations with periodic
boundary conditions highlighted an anomaly: fits to the leading corrections to
scaling for  the two estimators of the transition point described in the
previous section, $\beta^{C_V^{\rm max}}(L)$ and $\beta^{B^{\rm min}}(L)$, are
much better if a leading $1/L^2$ correction is assumed rather than the standard
$1/L^3$ \cite{us_goni}. The  goodness-of-fit parameter $Q$ for fits on the
extremal locations of the specific heat, $\beta^{C_V^{\rm max}}$, and Binder's
energy cumulant, $\beta^{B^{\rm min}}$, are shown in
Fig.~\ref{fig:fitting_quality}. Similar behaviour is seen in the dual model of
Eq.~(\ref{dual2}).

A resolution to the puzzle follows from noting that we have implicitly assumed
that the degeneracy $q$ of the low-temperature phase is fixed, and in
particular is independent of the system size in the calculations for the
heuristic model in the preceding section. However, for the  $3d$ plaquette
Ising model $q$ is size-dependent: $q=2^{3L}$. It is also macroscopic, since it
is increasing exponentially with the system size, although  sub-extensive since
it is increases as $\exp(L)$ rather than $\exp(L^3)$. 

Taking $q=2^{3L}$ we can rework the calculation of the leading scaling terms in
Eqs.~(\ref{eq:fss:beta:specheat}), (\ref{eq:fss:beta:binder}) to take account
of the size-dependence \cite{us_goni}, giving
\begin{eqnarray} \beta^{C_V^{\rm max}}(L) &=& \beta^\infty - \frac{\ln
	2^{3L}}{L^3\Delta\hat{e}} + {\cal O}\left( (\ln 2^{3L})^2 L^{-6} \right)
\label{eq:fss:beta:specheat2} \\
&=& \beta^\infty - \frac{3\ln 2}{L^{2}\Delta\hat{e}} + {\cal O}\left( L^{-4}
\right)\nonumber
\end{eqnarray} 
and
\begin{eqnarray} 
\beta^{B^{\rm min}}(L) &=& \beta^\infty - \frac{\ln(2^{3L}\hat{e}^2_{\rm
		o}/\hat{e}^2_{\rm d})}{L^3\Delta \hat{e}} + {\cal O}\left(
(\ln(2^{3L}\hat{e}^2_{\rm o}/\hat{e}^2_{\rm d}))^2 L^{-6}\right)
\label{eq:fss:beta:binder2} \\ &=&
\beta^\infty - \frac{3\ln2}{L^2\Delta \hat{e}}  - \frac{\ln(\hat{e}^2_{\rm
		o}/\hat{e}^2_{\rm d})}{L^3\Delta\hat{e}} + {\cal O}\left( L^{-4}\right)\nonumber
\end{eqnarray} 
for the estimators of the finite-size transition points.  Both show that the
leading contribution to the finite-size corrections is now $\propto L^{-2}$
rather than the customary $L^{-3}$ expected for periodic boundary conditions
and size independent degeneracy.  For the extremal values one finds 
\begin{equation} 
C_V^{\rm max}(L) = L^3\left( \frac{\beta^\infty \Delta\hat{e}}{2}\right)^2 +
{\cal O}(L)
\label{eq:fss:specheat} 
\end{equation} 
and 
\begin{equation} B^{\rm
  min}(L) = 1 - \frac{1}{12}\left( \frac{\hat{e}_{\rm o}}{\hat{e}_{\rm d}} +
  \frac{\hat{e}_{\rm d}}{\hat{e}_{\rm o}} \right)^2  + {\cal O}(L^{-2}) \; .
\label{eq:fss:binder} 
\end{equation}
The inverse temperature where both peaks of the energy probability density have
equal {weight}, $\beta^{\rm eqw}(L)$, has a behaviour that, to leading order,
coincides with the scaling of the location of the specific-heat maximum in
Eq.~(\ref{eq:fss:beta:specheat2}),
\begin{equation} 
\beta^{\rm eqw}(L) = \beta^\infty - \frac{3\ln 2}{L^{2}\Delta\hat{e}} + {\cal
	O}\left( L^{-4} \right) \; .  
\label{eq:fss:beta:eqw} 
\end{equation} 
The results of numerical simulations of the $3d$ plaquette Ising model such as
those shown in Fig. \ref{fig:fitting_quality}  provide convincing verification
of the non-standard $1/L^2$ leading scaling corrections due to the macroscopic
degeneracy.

\section{Simulational Considerations}

Since in Refs.~\cite{us_goni,us_fukinuke} we employed multicanonical
simulations giving a flat energy distribution,
reweighting techniques
\cite{reweighting,mucaweights} can be used to get an estimator of the energy
probability densities at different temperatures. $\beta^{\rm eqw}$ is  chosen
systematically to minimise 
\begin{equation} 
D^{\rm eqw}(\beta) = \left( \sum_{e < e_{\rm min}} p(e, \beta) - \sum_{e \geq
	e_{\rm min}} p(e, \beta) \right)^2 
\end{equation}
where the energy of the minimum between the two peaks, $e_{\rm min}$, is
determined beforehand to distinguish between the different phases.  The
location of the minimum, $\beta^{\rm eqw}$, is then used to calculate the
energy moments of the {ordered and disordered} phases, 
\begin{eqnarray} 
e^k_{\rm o}(L) = \sum_{e < e_{\rm min}} e^k\, p(e, \beta^{\rm eqw}) \Big/ \sum_{e < e_{\rm min}} p(e, \beta^{\rm eqw}),\nonumber\\
e^k_{\rm d}(L) =
\sum_{e \geq e_{\rm min}} e^k\, p(e, \beta^{\rm eqw}) \Big/\sum_{e \geq e_{\rm min}} p(e, \beta^{\rm eqw}),
\label{eq:e-eqw}
\end{eqnarray} 
where $e_{\rm o/d}(L) = e^1_{\rm o/d}(L)$ is the energy in the respective
phases, and their difference is an estimator of the latent heat $\Delta
e(L)=e_{\rm d}(L)-e_{\rm o}(L)$. Also, the second and first moments combine to
give the specific heat of the ordered and disordered phases, 
\begin{equation} 
C_{\rm o/d}(L) = \beta^2L^d\left( e^2_{\rm o/d}(L) - \left(e_{\rm o/d}(L)\right)^2\right).
\end{equation}
To find the inverse transition temperature where both phases have equal height
we minimise 
\begin{equation} 
D^{\rm eqh}(\beta) =  \left( \max_{e < e_{\rm min}}\{p(e, \beta)\} - \max_{e
	\geq e_{\rm min}}\{p(e, \beta)\} \right)^2, 
\end{equation} 
as a function of $\beta$.  The probability density $p(e, \beta^{\rm eqh})$
itself at $\beta^{\rm eqh}$ is also of interest since one can make use of it to
extract the reduced interface tension
\begin{equation} 
\sigma(L) = \frac{1}{2L^2} \ln \left( \frac { \max \{ p(e, \beta^{\rm eqh})
	\} } { \min \{ p(e, \beta^{\rm eqh}) \} } \right),  
\label{eq:interface-tension}
\end{equation} 
for periodic boundary conditions.

If we collect the various estimates for physical quantities using the correct,
modified leading $1/L^2$ scaling corrections for periodic boundaries and $1/L$
corrections for fixed boundaries we get consistent values across the original
plaquette Hamiltonian with both fixed and periodic boundary conditions and the
dual Hamiltonian with periodic boundaries as shown in
Table~\ref{tab:infinite_system_results} \cite{us_goni}.  We find an overall
consistent value for the inverse transition temperature of 
\begin{equation}
\beta^\infty = \betainfall \; .
\label{eq:b_final}
\end{equation}
We also find values for the interface tension of the original model and its
dual with periodic boundary conditions of $\sigma = 0.12037(18)$ and  $\sigma =
0.1214(13)$, respectively. The interface tension of the original model with
fixed boundary conditions is found to be much smaller, $\sigma = 0.0281(7)$.
\begin{table}[tpb] 
	\centering 
	\caption{
		Overview of resulting quantities of the infinite systems. } \vspace{3ex}
	\begin{tabular}{*{7}{l}} 
		\toprule \multicolumn{1}{c}{model} &
		\multicolumn{1}{c}{bc} & \multicolumn{1}{c}{$\beta^\infty$} &
		\multicolumn{1}{c}{$\hat{e}_{\rm o}$} & \multicolumn{1}{c}{$\hat{e}_{\rm d}$} & \multicolumn{1}{c}{$\Delta\hat{e}$} &
		\multicolumn{1}{c}{$\hat\sigma$}\\ 
		\midrule
		original  & \!\!\!\!periodic & \!\!\!\!0.551332(8) & $\!\!\!\!-1.468364(4)$ & $\!\!\!\!-0.617396(11)$ & \!\!\!\!0.850968(18) & {\!\!\!\!0.12037(18)} \\
		dual         & \!\!\!\!periodic & \!\!\!\!0.55143(7)  & $\!\!\!\!-1.37644(21)$ & $\!\!\!\!-0.88227(19)$  & \!\!\!\!0.49402(26)  & \!\!\!\!0.1214(13)\\ 
		&          & \multicolumn{2}{l}{[$\beta^\infty_{\rm dual} = 1.31328(12) $]}\\ 
		original & \!\!\!\!fixed    &  \!\!\!\!0.55138(5)  & $\!\!\!\!-1.4782(27) $ &$\!\!\!\! -0.790(4)  $ & \!\!\!\!0.694(4)  & \!\!\!\!0.0281(7) \\ 
		\bottomrule
	\end{tabular} \label{tab:infinite_system_results} 
\end{table} 

\section{Order Parameter(s)}

From Fig.~\ref{fig:sketch} it is clear that the standard magnetisation in the
ordered low-temperature phase will be zero, since there is no energy penalty
for flipping the sign of a plane of spins and spin configurations in the
low-temperature phase could contain arbitrary numbers of flipped planes of
spins with respect to a purely ferromagnetic state.  A possible alternative
order parameter emerged from the work of Suzuki, who had investigated an
anisotropic variant of the plaquette model many years ago  \cite{suzuki_old}.
He dubbed this the  fuki-nuke (``no-ceiling'' in Japanese) model, because the
horizontal ceiling plaquettes have zero coupling. In the  fuki-nuke model it
was possible to define an (anisotropic) order parameter by using its hidden
equivalence to a stack of standard $2d$ Ising models.  Suzuki and collaborators
returned to plaquette Ising models much more recently  \cite{suzuki1} to
observe that a similar order parameter to that used for the fuki-nuke model
might still be viable for the isotropic $3d$ plaquette Ising Hamiltonian, but
in that case it should be independent of the orientation.

The anisotropic $3d$ plaquette Hamiltonian for the fuki-nuke model is given by
\bea
H_{\rm fuki\mhyphen nuke} &=& - J_x  \sum_{x=1}^{L} \sum\limits_{y=1}^{L}\sum\limits_{z=1}^{L} \sigma_{x,y,z} \sigma_{x,y+1,z}\sigma_{x,y+1,z+1} \sigma_{x,y,z+1} \nonumber \\
&{}&  - J_y \sum_{x=1}^{L} \sum\limits_{y=1}^{L}\sum\limits_{z=1}^{L}  \sigma_{x,y,z} \sigma_{x+1,y,z}\sigma_{x+1,y,z+1} \sigma_{x,y,z+1}\;,
\eea
where we have now written the individual spin positions explicitly and set the
coupling of the horizontal plaquettes $J_z=0$. This Hamiltonian may be
rewritten as a stack of standard $2d$ nearest-neighbour Ising models by
defining bond spin variables $\tau_{x,y,z} = \sigma_{x,y,z} \sigma_{x,y,z+1}$
at each vertical lattice bond. With periodic (vertical) boundary conditions we
must impose the constraints $\prod_{z=1}^L\tau_{x,y,z} = 1$ which preclude an
explicit solution \cite{us_fuki}, but in the case of free boundary conditions 
the Ising layers completely decouple, with the Hamiltonian
\be
H_{\rm fuki\mhyphen nuke} = - \sum\limits_{x=1}^{L}\sum\limits_{y=1}^{L}\sum\limits_{z=1}^{L} \left( \tau_{x,y,z}  \tau_{x+1,y,z} + \tau_{x,y,z} \tau_{x,y+1,z} \right)\;,
\label{stack}
\ee
where we have set $J_x=J_y=1$ for simplicity. The partition function for free
boundary conditions is
 
\begin{eqnarray}
Z_{\rm fuki\mhyphen nuke} &=& \sum_{\{\tau\}} \exp\left(-\beta H_{\rm fuki\mhyphen nuke}(\{\tau\})\right)\nonumber\\
&=& 2^{L^2}\sum_{\{\tau_{x,y,z\neq L}\}}\prod_{z=1}^{L-1}\exp\left(\beta \sum\limits_{x=1}^{L}\sum\limits_{y=1}^{L}\left( \tau_{x,y,z}  \tau_{x+1,y,z} + \tau_{x,y,z} \tau_{x,y+1,z} \right)\right)\nonumber\\
&=& 2^{L^2}\prod_{z=1}^{L-1}\sum_{\{\tau_{x,y}\}_z}\exp\left(\beta \sum\limits_{x=1}^{L}\sum\limits_{y=1}^{L}\left( \tau_{x,y,z}  \tau_{x+1,y,z} + \tau_{x,y,z} \tau_{x,y+1,z} \right)\right)\nonumber\\
&=& 2^{L^2}\prod_{z=1}^{L-1} Z _{2d\;\rm Ising} = 2^{L^2} \left(Z_{2d\;\rm Ising}\right)^{L-1},
\label{eq:fukinuke-free}
\end{eqnarray}
where $\{\tau_{x,y}\}_z$ denotes summation over all $\tau$-spins with a given
$z$-component and $Z_{2d\;\rm Ising}$ is the standard $2d$ Ising model
partition function.
Each $2d$ Ising layer in Eq.~(\ref{eq:fukinuke-free}) magnetises independently
at the $2d$ Ising model transition temperature. For the case of periodic
boundaries we would expect the same thermodynamic behaviour with only the
corrections to scaling being affected by the constraints arising from the
boundary conditions. 

When the model is expressed in terms of the $\tau$ spins, a suitable order
parameter for a single Ising layer is the standard magnetisation 
\be
m_{2d, z} =  \left< \frac{1}{L^2} \sum_{x=1}^{L} \sum\limits_{y=1}^{L} \tau_{x,y,z} \right>
\label{Mone}
\ee
which when translated back to the original $\sigma$ spins gives
\be 
m_{2d, z} = \left<  \frac{1}{L^2} \sum_{x=1}^{L} \sum\limits_{y=1}^{L} \sigma_{x,y,z} \sigma_{x,y,z+1}   \right> 
\ee
This single layer nearest (vertical) neighbour correlator in the fuki-nuke model
will thus behave like the standard $2d$ Ising  magnetisation 
$\pm \left| \beta - \beta_c \right|^{1 \over 8}$ near the critical point
$\beta_c = \frac{1}{2}\ln ( 1 + \sqrt{2})$.  

There are two obvious possibilities for constructing a pseudo-$3d$ order
parameter from the layer magnetisations in the fuki-nuke model, both chosen to
eliminate cancellations between differently magnetised planes of Ising spins
\cite{me_fukinuke}.  The first is to take the absolute value of the
magnetisation in each plane
\be
m_{\rm abs} = \left< \frac{1}{L^3}\sum_{z=1}^{L} \left| \sum\limits_{x=1}^{L}\sum\limits_{y=1}^{L} \sigma_{x,y,z}\sigma_{x,y,z+1}\right| \right>\;,
\label{Mabs}
\ee
and the second is to square the magnetisation of each plane,
\be
m_{\rm sq} =    \left< \frac{1}{L^5} \sum_{z=1}^{L}  \left( \sum\limits_{x=1}^{L}\sum\limits_{y=1}^{L}   \sigma_{x,y,z}\sigma_{x,y,z+1}  \right)^2  \right> \;,
\label{Msq}
\ee
We have retained the various normalizing factors in Eqs.~(\ref{Mabs}) and
(\ref{Msq}) for a cubic  lattice with $L^3$ sites. 

Suzuki's and Hashizume's suggestion was that we could define the order parameter for
the isotropically coupled plaquette model in a similar fashion by using the
same nearest-neighbour correlator in any of the three directions, e.g.,
\begin{eqnarray}
m_{\rm abs}^x = \left< \frac{1}{L^3}\sum_{x=1}^{L} \left|\sum\limits_{y=1}^{L}\sum\limits_{z=1}^{L} \sigma_{x,y,z}\sigma_{x+1,y,z}\right| \right>\;,
\end{eqnarray}
with the obvious analogous definition for the $y,z$ directions and where we
have again assumed periodic boundary conditions.  The squared magnetisations
could be defined in a similar manner as
\begin{eqnarray}
m_{\rm sq}^x = \left< \frac{1}{L^5}\sum_{x=1}^{L} \left(\sum\limits_{y=1}^{L}\sum\limits_{z=1}^{L} \sigma_{x,y,z}\sigma_{x+1,y,z}\right)^2 \right>\;, \\
\nonumber
\end{eqnarray}
and similarly for the $y,z$ directions. In the isotropic case we would expect
$m_{\rm abs}^x = m_{\rm abs}^y = m_{\rm abs}^z$ and similarly for the squared
quantities.  Lipowski had previously suggested \cite{Lip1} using an order
parameter akin to $m_{\rm sq}$ in Eq.~(\ref{Msq}).

\begin{figure}[h!]
	\begin{center} 
		\includegraphics[width=0.5\textwidth]{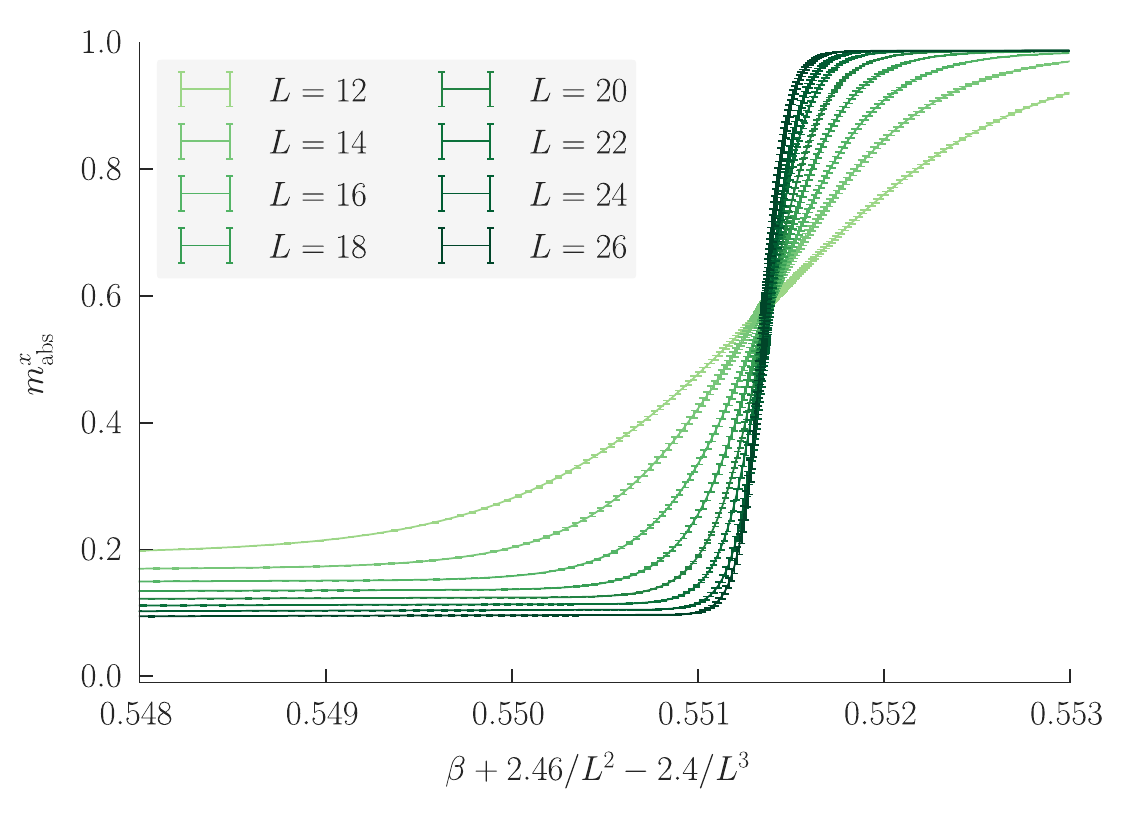}\hfill
		\includegraphics[width=0.5\textwidth]{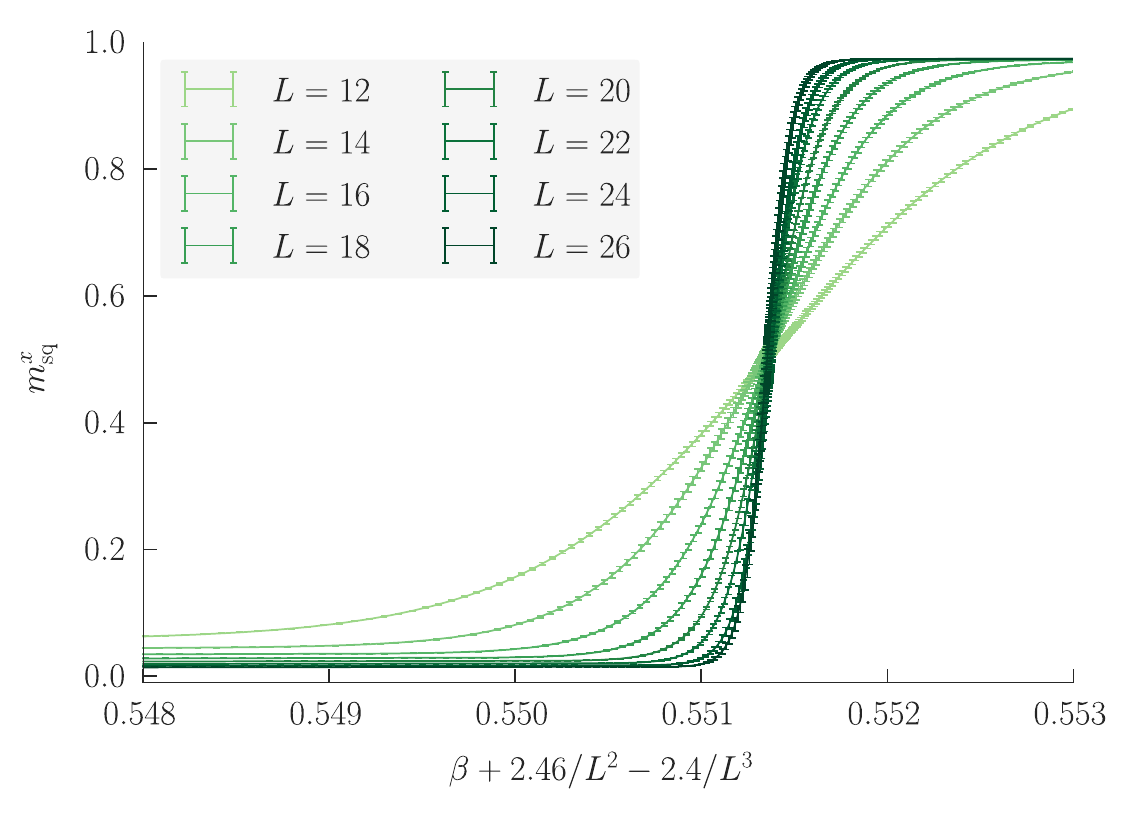}\\
		\includegraphics[width=0.5\textwidth]{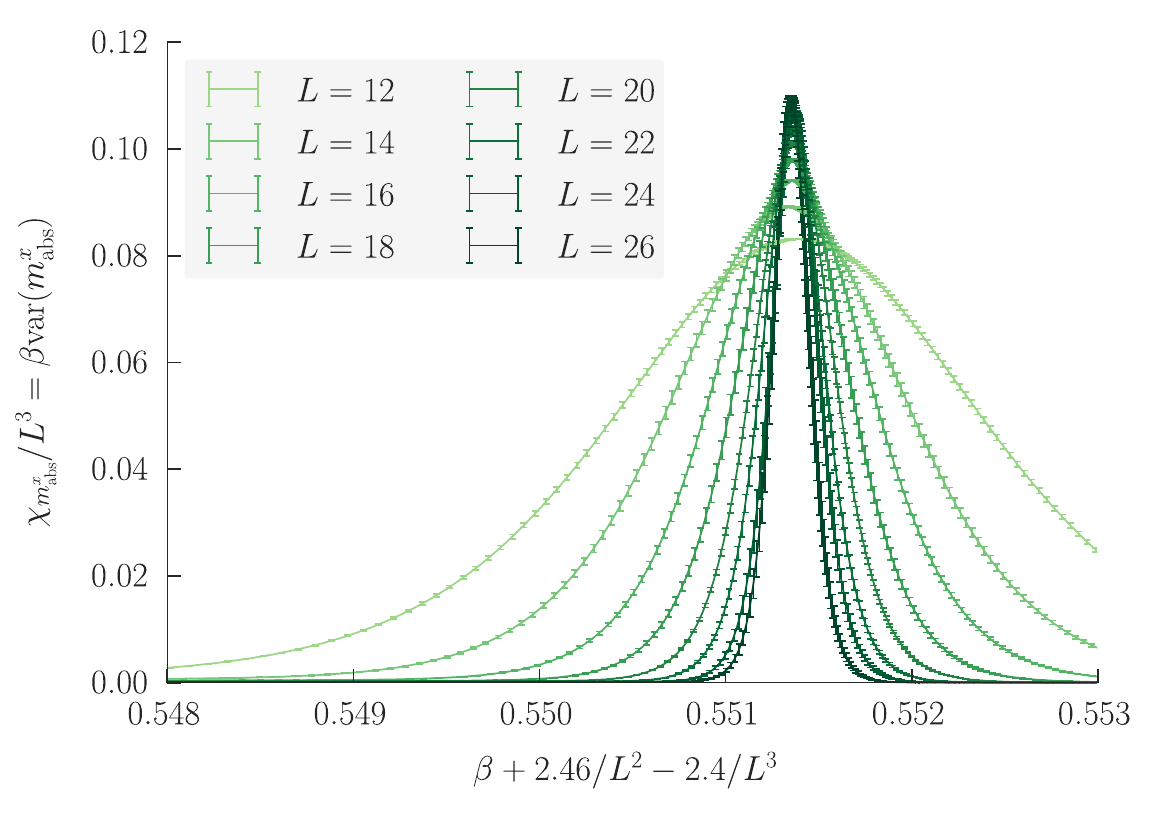}\hfill
		\includegraphics[width=0.5\textwidth]{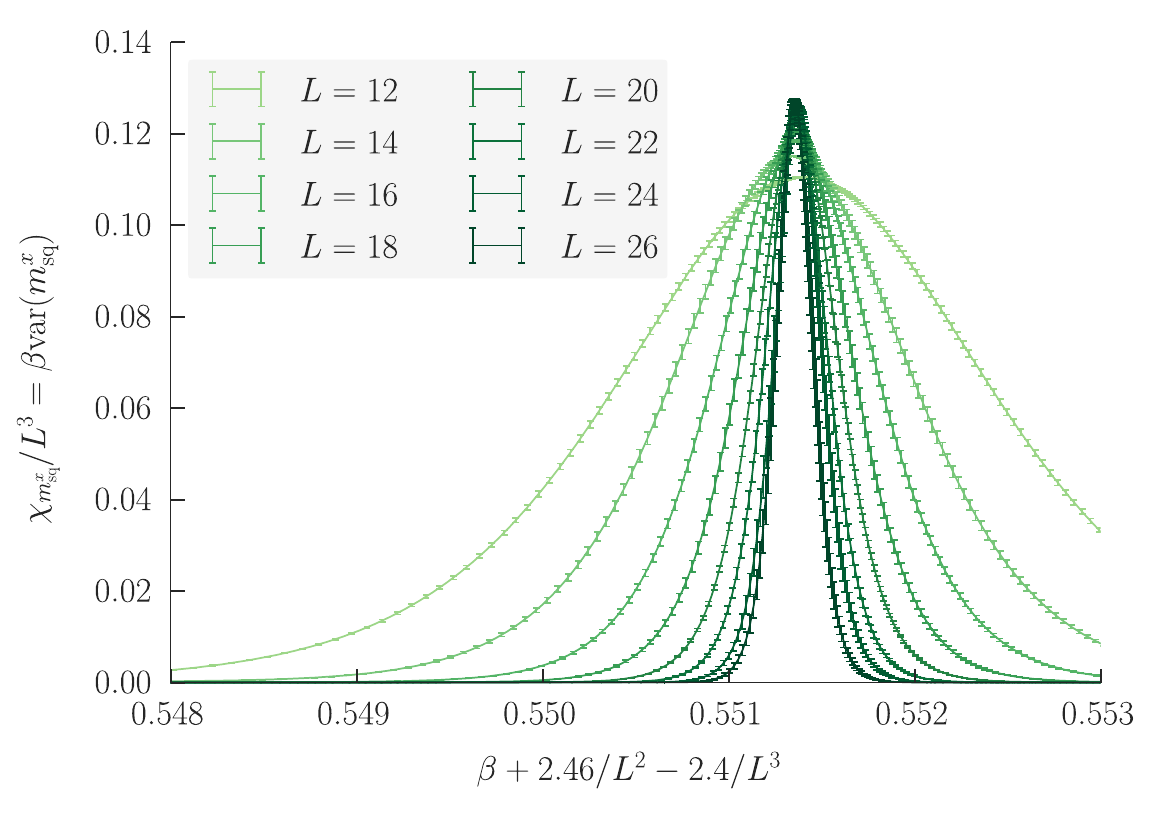}
		\caption{The fuki-nuke parameters $m^{x}_{\rm abs}$
			and $m^{x}_{\rm sq}$ (upper row) along with their respective
			susceptibilities $\chi$ normalised by the system volume (lower row) over 
			shifted inverse temperature $\beta$ for
			several lattice sizes $L$.}
			\label{fig:canonmx} 
	\end{center} 
\end{figure}

If we consider the susceptibilities $\chi$
for the various 
order-parameter candidates, the peak locations $\beta^{\chi}(L)$ 
for the different lattice sizes $L$
can be fitted using the modified first-order scaling laws discussed in the
preceding section,
\begin{eqnarray} \beta^{\chi}(L) = \beta^{\infty} + a/L^2 +
  b/L^3\;.
\end{eqnarray} 
This gives for the estimate of the inverse critical
temperature $\beta^{\chi}(L)$ from the fuki-nuke susceptibility $\chi_{m_{\rm
abs}^{x}}$
\begin{eqnarray}
	\beta^{\chi_{m_{\rm abs}^{x}}}
	(L) = 0.551\,37(3) - 2.46(3)/L^2 + 2.4(3)/L^3\;,
\end{eqnarray}
with a goodness-of-fit parameter $Q=0.64$ and $12$ degrees of freedom left as
shown in Fig.~\ref{fig:canonmx}.  Fits to the other directions $m_{\rm
abs}^{y,z}$ and fits to the peak location of the susceptibilities of $m_{\rm
sq}^{x,y,z}$ give  the same parameters within error bars and are of comparable
quality.
The estimates of the transition point obtained from the various $m_{\rm abs}$
and $m_{\rm sq}$ are thus independent of the directions $x,y,z$, confirming
Suzuki's and Hashizume's hypothesis, and their values are consistent with the 
estimates in Table~\ref{tab:infinite_system_results} and Eq.~(\ref{eq:b_final})
obtained from $\beta^{C_V^{\rm max}}(L)$ and $\beta^{B^{\rm min}}(L)$
\cite{us_fukinuke}.

We can see that the low-temperature degeneracy has impacted both the definition
of the magnetic order parameter, giving rise to an order parameter that is a
hybrid $2d/3d$ construct and the  finite-size scaling  of the associated
susceptibility, which also shows the leading $1/L^2$ corrections seen in
energetic quantities such as the specific-heat peak. 

\section{A Curious Symmetry -- Quantum Aspects: Duality and Fractons}
A further consequence of the planar flip symmetry is found in a Hamiltonian
related to the quantum dual of the plaquette model. This fits into the general
framework developed in \cite{fracton1,fracton2} in which novel fracton
topological phases are constructed by gauging  symmetries acting on subsystems
of dimension $2 \le d_{s} < d$. Since the spin-flip symmetry in the $3d$
plaquette model acts on $2d$ planes it has precisely this property. The
procedure for constructing the fracton Hamiltonian follows closely that of the
Kitaev toric code, giving commuting electric and magnetic operators. 

The quantum spin version of the plaquette model promotes the classical Ising
spins $\sigma$ to operators, represented by Pauli matrices $\sigma^z$ and spin
flips are implemented with $\sigma^{x}$,%
\footnote{At the risk of causing confusion we have retained our nomenclature
  $\sigma$ for the original spins and called the dual nexus spins $\tau$, which
  is the opposite of the convention employed in \cite{fracton1}.} so an
  external transverse field term may be added to the Hamiltonian  to give
\begin{equation}
H_0 =  -  t \sum_{[i,j,k,l]}\sigma^z_{i} \sigma^z_{j}\sigma^z_{k} \sigma^z_{l} - h\sum_{i}\sigma^{x}_{i}\;.
\end{equation}
The quantum dual of this Hamiltonian is given by
\begin{equation}
H_{\rm nexus} = -  t \sum_i \tau^z_i - h \sum_i A_i \;,
\end{equation}
where the nexus spin operators $\tau$ live at the centre of each
plaquette on the original lattice, i.e., the centre of links on the dual
lattice.  They play the same role as the link gauge spins introduced in
dualising the standard $\mathbb{Z}_2$ Ising model in $3d$.  This generalised
gauging/dualising procedure may be applied to other classical systems with a
subsystem symmetry and leads, inter alia, to the Haah code model \cite{fracton_hist}, 
as well as other fracton models.

With the nexus spins in $H_{\rm nexus}$ residing at the centres of the links of
the dual cubic lattice, the  nexus charge operator $A_i$ is given by
\begin{equation}
A_{i} \equiv \prod_{j \in P(i)}\tau^{x}_{j} \;,
\end{equation}
where the  set $P(i)$ specifies the locations of multi-spin interactions that
are affected by a spin flip using $\sigma^{x}_{i}$ in $H_0$.  For the plaquette
model $P(i)$ comprises the twelve $\tau$ spins  on the edges of a cube in the
dual lattice as shown in Fig.~6, coming from the twelve plaquettes that are
affected by acting with a single spin-flip operator $\sigma^{x}_{i}$  in $H_0$,
which lies at the centre of the dual lattice cube.

\begin{figure}[t]
	\label{Xcube} 
	\begin{center}	
		\includegraphics[height=4cm]{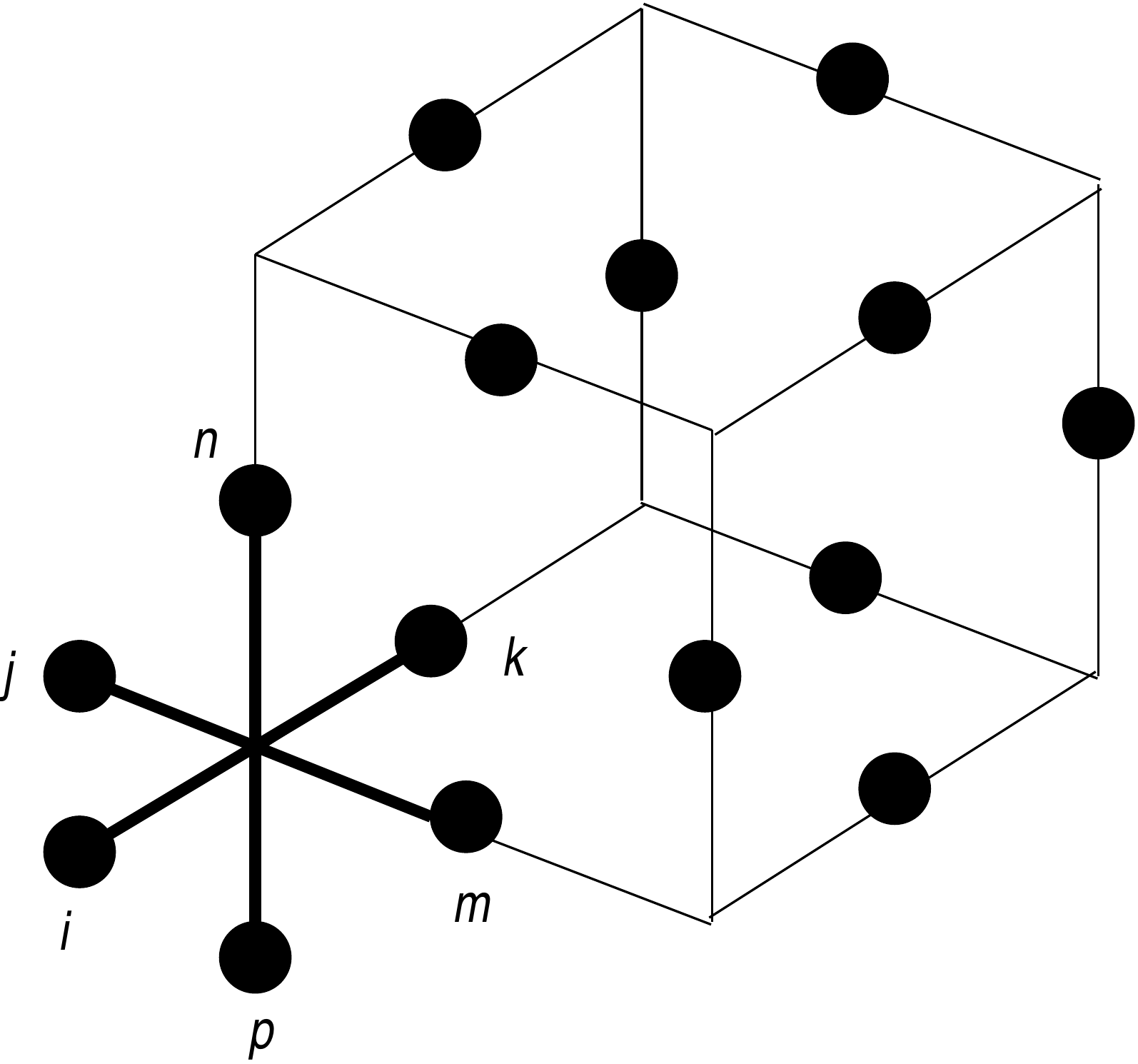}
    \caption{The spins involved in defining the fracton Hamiltonian, 
      $H_{\rm fracton}$, shown on the edges of a dual lattice cube (for
      $A_i$ terms) and on a ``star'' at one corner (for $B_i$ terms).}
	\end{center}
\end{figure}

We also need to impose constraints on the physical states in the theory that
arise from rewriting $H_0$ in terms of the nexus spins, $\tau$, as is apparent
already at the classical level. If we consider the product of 
\begin{figure}[b]
	\label{matchbox1} 
	\begin{center}	
		\includegraphics[height=3cm]{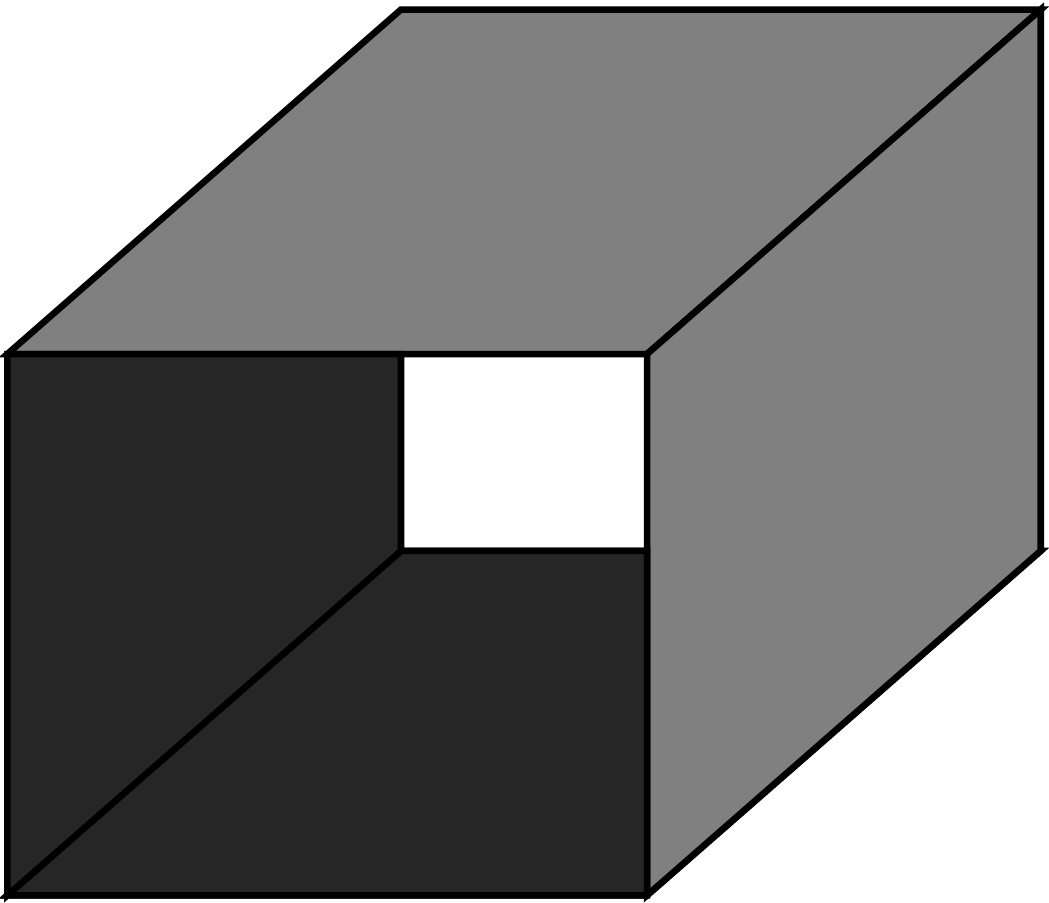}
    \caption{One of the three possible orientations of a product of four
      plaquettes round a cube.}
	\end{center}
\end{figure}
four plaquettes around a cube shown in Fig.~\ref{matchbox1} we can see that
this must be one, and similarly for the other orientations.  This means that
products of the nexus spins associated with  the shaded plaquettes must also be
one and translates into constraints on the physical states in the quantum spin
version of the theory,
\begin{equation}\label{eq:Flatness}
  B_{i}\ket{\Psi} = \ket{\Psi} \;,
\end{equation}
where the $B_{i}$ are the three star products of the nexus spin operators 
\begin{eqnarray}
  B_{i}^{(xz)} &=& \tau^z_i \tau^z_p \tau^z_n \tau^z_k \;,\nonumber \\
  B_{i}^{(xy)} &=& \tau^z_i \tau^z_j \tau^z_k \tau^z_m \;, \\
  B_{i}^{(yz)} &=& \tau^z_j \tau^z_n \tau^z_m \tau^z_p \nonumber
\end{eqnarray}
and can also be thought of as generalised monopole charge operators.  These are
also shown in Fig.~6.  The effect of the constraints is to restrict the Hilbert
space of the $\{ \tau \}$ to be the domain wall configurations of the ordered
phase of $H_{0}$.  It is interesting to note that a different approach in which
dual spins are placed at the centre of the cubes wrapped by the shaded
plaquettes, rather than at the centre of each plaquette,  leads to the
Ashkin-Teller-like Hamiltonian of Eq.~(\ref{dual2}).

The X-Cube fracton Hamiltonian
\begin{equation}
H_{\rm fracton} =  - \sum_i B_i - \sum_i A_i
\end{equation}
named after the configuration of the interaction terms seen in Fig.~6 was
argued in \cite{fracton1} to display fracton topological order with two
distinguishing characteristics: it had a  sub-extensive topological degeneracy
and the motion of both electric and magnetic excitations was constrained. 

This flip symmetry for planes of spins plays a central role in both these
deductions. In the original plaquette Hamiltonian $H_0$ the symmetry was
generated by products of $\sigma_i^x$ in the plane. As we have seen, the
associated ground states in the classical system have a sub-extensive $2^{3L}$
degeneracy, which remains the case in the quantum system when $t/h \gg 1$. The
dual representation of the planar product of $\sigma_i^x$'s is a product of the
nexus charges over the same plane and the plaquette interaction terms in $H_0$
are dual to individual nexus spins $\tau_i^z$.  The dual relation to the spin
flip commuting with the plaquette interaction term is thus
\begin{equation}
\left[
\tau_i^z, \prod_{i \in \Sigma} A_i 
\right]  = 0
\end{equation} 
where $\Sigma$ is a planar region along which spins are flipped.  This can only
be satisfied if $ \prod_{i \in \Sigma} A_i=1$ on such a plane, so each plane
provides one constraint for the nexus charges.

In general with $N$ nexus spins $\tau$ and $M$ nexus charges $A_i$ and
monopole operators $B_i$ in a theory, the topological ground-state degeneracy
on a torus (i.e., a lattice with periodic boundary conditions) would be $D=2^{k +
M-N}$ where $k$ is the number of constraints satisfied by the $A_i$ and $B_i$.
For the fracton Hamiltonian $M=N$ and the number of constraints is $k=3L$, one
from each plane, from the arguments above. The sub-extensive ground-state
degeneracy due to the  flip symmetry of the plaquette model thus translates
into the sub-extensive topological degeneracy of the fracton Hamiltonian.

A second consequence of the symmetry is that the excitations in the X-Cube
model have restricted mobility. The location of electric excitations created in
the ground state of $H_{\rm fracton}$ by a product of $\tau_i^z$ operators on
some planar subset $\Sigma$ of the lattice
\begin{equation}
W = \prod_{i \in \Sigma} \tau^z_i
\end{equation}
is identical to the location of the planar spin flips created by the dual
operator
\begin{equation}
\tilde W = \prod_{\Box_i} \sigma^z \sigma^z \sigma^z \sigma^z
\end{equation}
acting on the paramagnetic state of $H_0$, where the product is over the
plaquettes $\Box_i$ associated with the nexus spins $\tau_i^z$ in the same
region $\Sigma$.

Since $\tilde W$ cannot create an isolated pair of spin-flip excitations in the
paramagnetic state, $W$ cannot create an isolated pair of fracton excitations
and four will appear at the corners of a rectangular region of flipped spins.
This is characteristic of so-called type I fracton order \cite{fracton1}.
Although the fracton (electric) excitations are pinned in this model, a Wilson
line of $\tau^x$ operators acting on the fracton Hamiltonian ground state
generates a pair of magnetic excitations (which come in two flavours) at the
ends of the line which can continue to  move along the line. Changing direction
generates a further magnetic excitation of a different flavour at the corner.

\section{Conclusions}

The $3d$ plaquette Ising model possesses an unusual planar spin-flip symmetry
and macroscopic (but sub-extensive) low-temperature phase degeneracy.  This has
consequences for the finite-size scaling at its first-order transition, which
displays $1/L^2$ leading corrections, and for the definition of the magnetic
order parameter, which has a hybrid $2d/3d$ form. We described the symmetry in
the plaquette model and its dual and outlined the necessary adaptions to
standard first-order finite-size scaling in order to take account of
macroscopic degeneracy. The anisotropic, fuki-nuke, variant of the plaquette
model was then used to deduce a  suitable order parameter.

Finally, we looked at implications of the symmetry in the quantum version of
the model. Here, a  Hamiltonian related to the quantum dual of the plaquette
model describes fracton topological phases, which have point-like topological
excitations that appear at the corners of membrane-like operators, such as
planar regions of a $3d$ lattice. This fracton Hamiltonian displays  a
macroscopic, sub-extensive topological degeneracy that is  directly related to
the spin-flip symmetry of the plaquette model, as is the restricted mobility of
the topological excitations in the model.

Further numerical exploration of models with subsystem symmetries analogous to the
plaquette model might prove
profitable for the understanding of  exotic topological phases of matter, such
as the fracton phase discussed here. For instance, Monte Carlo simulations
of the ($3d$, classical) ${\mathbb Z}_2$ gauge-Higgs model were employed to good
effect  in investigating the phase diagram of the ($2d$, quantum) toric code by using
the $d$-dimensional quantum to $d+1$-dimensional classical mapping \cite{ToricHiggs}.

\begin{acknowledgement}
We would like to thank the organizers of the PTCP conference.  This work was
supported by the Deutsche Forschungsgemeinschaft (DFG) through the
Collaborative Research Centre SFB/TRR 102 (project B04), the
Deutsch-Franz\"osische Hochschule (DFH-UFA) through the Doctoral College
``$\mathbb{L}^4$'' under Grant No.\ CDFA-02-07, and by the EU Marie Curie IRSES
Network DIONICOS under Contract No. PIRSES-GA-2013-612\,707.
\end{acknowledgement}

\end{document}